\documentclass[twocolumn,showpacs,preprintnumbers,amsmath,amssymb,pra]{revtex4}

\usepackage{graphicx}
\usepackage{dcolumn}
\usepackage{bm}

\begin{document}

\title{From Optical Lattice Clocks to the Measurement of Forces in the Casimir Regime}

\author{Peter Wolf$^{1,2}$}
\email{peter.wolf@obspm.fr}
\author{Pierre Lemonde$^1$}
\author{Astrid Lambrecht$^3$}
\author{S\'ebastien Bize$^1$}
\author{Arnaud Landragin$^1$}
\author{Andr\'e Clairon$^1$}
\affiliation{$^1$LNE-SYRTE, Observatoire de
Paris\\ 61, Avenue de l'observatoire, 75014, Paris, France\\ $^2$Bureau International des Poids et Mesures, 92312 S{\`e}vres Cedex, France\\ $^{3}$LKB, Universit\'e Pierre et Marie Curie, Campus Jussieu, 75252 Paris,
France}

\date{\today}

\begin{abstract}
We describe a novel experiment based on atoms trapped close to a macroscopic surface, to study the interactions between the atoms and the surface at very small separations (0.6 to 10~$\mu$m). In this range the dominant potential is the QED interaction (Casimir-Polder and Van der Waals) between the surface and the atom. Additionally, several theoretical models  suggest the possibility of Yukawa type potentials with sub-mm range, arising from new physics related to gravity. The proposes set-up is very similar to neutral atom optical lattice clocks, but with the atoms trapped in lattice sites close to the reflecting mirror. A sequence of pulses of the probe laser at different frequencies is then used to create an interferometer with a coherent superposition between atomic states at different distances from the mirror (in different lattice sites). Assuming atom interferometry state of the art measurement of the phase difference and a duration of the superposition of about 0.1~s we expect to be able to measure the potential difference between separated states with an uncertainty of $\approx 10^{-4}$~Hz. An analysis of systematic effects for different atoms and surfaces indicates no fundamentally limiting effect at the same level of uncertainty, but does influence the choice of atom and surface material. Based on those estimates, we expect that such an experiment would improve the best existing measurements of the atom-wall QED interaction by $\geq$~2 orders of magnitude, whilst gaining up to 4 orders of magnitude on the best present limits on new interactions in the range between 100~nm and 100~$\mu$m.
\end{abstract}

\maketitle

\section{Introduction} \label{intro}

One of the frontiers of modern physics is the study of forces at very small length scales ($<$\,1\,mm). From the theoretical point of view, such length scales are the domain where QED interactions (Casimir type forces) become important and where several recent theoretical ideas point to the possibility of new interactions related to gravity (\cite{Dimopoulos,Fischbach} and references therein). On the experimental side, measurements at distances ranging from $10^{-8}$\,m to $10^{-3}$\,m have been the domain of microelectromechanical systems (MEMS) and of torsion balance experiments (\cite{Fischbach,FischbachPRD,Hoyle,Dimopoulos} and references therein). Two major difficulties of such mechanical experiments is the exact knowledge of the geometry of the setup (distance, surface roughness, etc...) and the precise measurement of the very small forces involved. A promising alternative that might provide a way around those difficulties is the use of cold atoms.

Over the last decade or so this idea has lead to a number of proposals and experiments that explore interactions at very short range using the metrological advantages of cold atoms and Bose Einstein condensates (BEC). The experiments that have been carried out so far \cite{Sukenik,Arnaud,Shimizu,Druzhinina,Lin,Pasquini,Oberst,Cornell} all confirm the theoretical predictions from QED (Van der Waals and Casimir-Polder effect) at distances ranging from a few tens of nanometers to several microns, however, none of them have yet reached the uncertainties achieved by the best mechanical measurements.

A promising technique to improve on those results is the use of cold atoms trapped in a standing wave close to a macroscopic surface as proposed in \cite{Dimopoulos,Inguscio,Tino}. In these schemes BECs or cold atoms are trapped coherently in several potential wells of the standing wave close to the surface and released after an interaction time $T$. The differing potential in the wells as a function of distance from the surface leads to a phase difference cumulated during $T$. This phase difference is then observed via imaging of the interference pattern once the trap is switched off and the BECs expand \cite{Dimopoulos}, or more realistically by observing the period of Bloch oscillations induced by the presence of gravity \cite{Inguscio,Tino}. A similar scheme has already been used for one of the most accurate determinations of the fine structure constant via the measurement of $\hbar/m_{Rb}$ that governs the period of Bloch oscillations \cite{Biraben2,Biraben}.

In this paper we pursue a similar idea (atoms trapped in a standing wave close to the surface) but propose a new scheme that has the advantage of providing accurate control of the position of the atoms and accurate detection of the cumulated phase difference, which in turn allows for a sensitive determination of the atom-surface separation and of the potential difference between the wells. This is achieved by a sequence of laser pulses at different frequencies that allow the creation of an interferometer with separated spatial paths (passing through different wells) and using two different internal states of the atom. The readout of the phase difference is then simply the detection of the internal state populations (similar to atomic clocks). The involved technology is very similar to that of existing setups for optical lattice clocks \cite{Katori,Ye2,Pierre2,Barber,Pierre} that have been built over the last years and are now operating in several laboratories around the world, which makes this a promising idea for the near future as experiments can draw from that experience. Most of those clocks are operating using $^{87}$Sr at a lattice laser wavelength of $813$\,nm and all numerical values throughout the paper are given for that case, except when explicitly stated otherwise.

We first recall (section \ref{WS}) the relevant results of a previous paper \cite{LW} in which we have studied the use of Wannier-Stark (WS) states in an accelerated periodic potential to reduce the required trap depth in optical lattice clocks, followed by their application to the control of atoms in an optical lattice in the presence of gravity (section \ref{control}). We then describe the proposed experiment and its potential statistical uncertainty in section \ref{experiment}. With that in mind we investigate potential perturbing effects (section \ref{systematics}) and finally study the resulting interest of the measurement in fundamental physics in sections \ref{QED} and \ref{YukS}. 

\section{General Theory} \label{WS}

We consider two level atoms trapped in a vertical standing wave with wave vector $k_l$ in the presence of gravity. The trapping laser is red-detuned from resonance which leads to trapping of the atoms at the maxima of intensity. Transverse confinement is provided by the Gaussian profile of the vertical laser or by a 3D lattice, adding two horizontal standing waves.

The internal atomic structure is approximated by a two-level
system $|g\rangle$ and $|e\rangle$ with energy difference
$\hbar\omega_{eg}$. The internal Hamiltonian is:
\begin{equation}
    \hat{H}_{i}=\hbar\omega_{eg}|e\rangle \langle e|.
\end{equation}
For $^{87}$Sr the transition used is the $^1S_0-^3P_0$ line at 698\,nm.

We introduce the coupling between $|e\rangle$ and $|g\rangle$ by a
laser (probe laser) of frequency $\omega$, initial phase $\phi$ and wave vector $k_s$ propagating along
the $x$ direction:
\begin{equation}
    \hat{H}_{s}=\hbar\Omega \cos(\omega t-k_s\hat{x}+\phi_s)|e\rangle \langle
    g|+h.c.,
\end{equation}
with $\Omega$ the Rabi frequency.

In the following we consider external potentials induced by trap
lasers and gravity. The
external potential $\hat{H}_{ext}$ is then identical for both
$|g\rangle$ and $|e\rangle$ with eigenstates $|m\rangle$ obeying
$\hat{H}_{ext}|m\rangle=\hbar \omega_m | m\rangle$. If we
restrict ourselves to experiments much shorter than the lifetime
of state $|e\rangle$ (for $^{87}$Sr, the lifetime of the lowest
$^3P_0$ state is about 100 s) spontaneous emission can be neglected and
the evolution of the general atomic state
\begin{equation}
    |\psi_{at}\rangle = \sum_m a_m^g \,e^{-i\omega_m t}\,|m,g\rangle + a_m^e
\,e^{-i(\omega_{eg}+\omega_m)t}\,|m,e\rangle
\end{equation}
is driven by
\begin{equation}
    i\hbar \frac{\partial}{\partial t}
    |\psi_{at}\rangle=(\hat{H}_{ext}+\hat{H}_{i}+\hat{H}_{s})|\psi_{at}\rangle .
\end{equation}

For a vertical optical lattice in the presence of gravity the external Hamiltonian is
\begin{equation} \label{Hext}
    \hat{H}_{ext}=\frac{\hbar^2 \hat{\kappa}^2}{2
m_{a}}+\frac{U_0}{2}(1-\cos(2k_l\hat{x}))+m_ag\hat{x},
\end{equation}
with $\hbar {\kappa}$ and $m_{a}$ the atomic momentum and mass, $U_0$ the depth of the trapping potential and $g$ the acceleration of the Earth's gravity. The natural energy unit for
the trap depth is the recoil energy associated with the
absorption or emission of a photon of the lattice laser,
$E_r=\frac{\hbar^2k_l^2}{2m_a}$ and in the following we consider values of $U_0$ ranging from 5\,$E_r$ to 100\,$E_r$.

The Hamiltonian (\ref{Hext}) supports no true bound states, as an atom initially confined in
one well of the lattice will end up in the continuum due to
tunneling under the influence of gravity, an effect known as Landau-Zener
tunneling. However, in the case of Sr in
an optical lattice, and for $U_0$ as low as $5\,E_r$ the timescale for this effect is about $10^{10}$\,s for atoms in the lowest lying state, so it can be safely neglected for our purposes.

\begin{figure}[b]
\begin{center}
\includegraphics[width=8.5 cm]{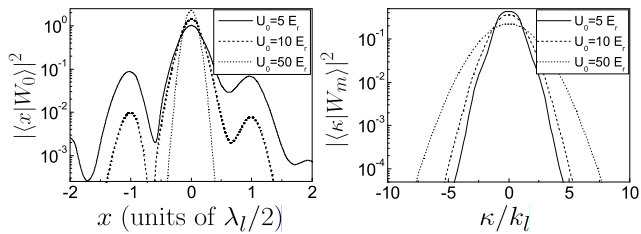}
\end{center}
\caption{Wannier-Stark states in position (left) and momentum
(right) representation for $U_0=5\,E_r$, $U_0=10\,E_r$ and
$U_0=50\,E_r$, calculated numerically (see \cite{LW} for details).}\label{WS_states}
\end{figure}

As shown in \cite{LW} under these conditions the Eigenstates of $\hat{H}_{ext}$ are the so called Wannier-Stark (WS) states $|W_m\rangle$ known from solid state physics \cite{Nenciu}. In the position representation
$|W_m\rangle$ exhibits a main peak in the $m^{\textrm{th}}$ well
of the lattice and small revivals in adjacent wells (see figure \ref{WS_states}). These
revivals decrease exponentially at increasing lattice depth. At
$U_0=10\,E_r$ the first revival is already a hundred times smaller
than the main peak, which indicates strong localization in the $m^{\textrm{th}}$ well. The discrete
quantum number $m$ is the "well index" characterizing the well
containing the main peak of the wave function $\langle
x|W_{m}\rangle$. As intuitively expected, the energy
separation between adjacent states is simply the change in
gravitational potential between adjacent wells: $\hbar
\Delta_g=m_ag\lambda_l/2$ (see \cite{LW} for details, for our case $\Delta_g/2\pi \approx$\,866\,Hz), which leads to a Wannier Stark ladder of Eigenstates as shown on figure \ref{fig:wannier_stark_couplings}.

\section{Control of Atoms in WS states} \label{control}

\begin{figure}[b]
\begin{center}
  \includegraphics[width=8 cm]{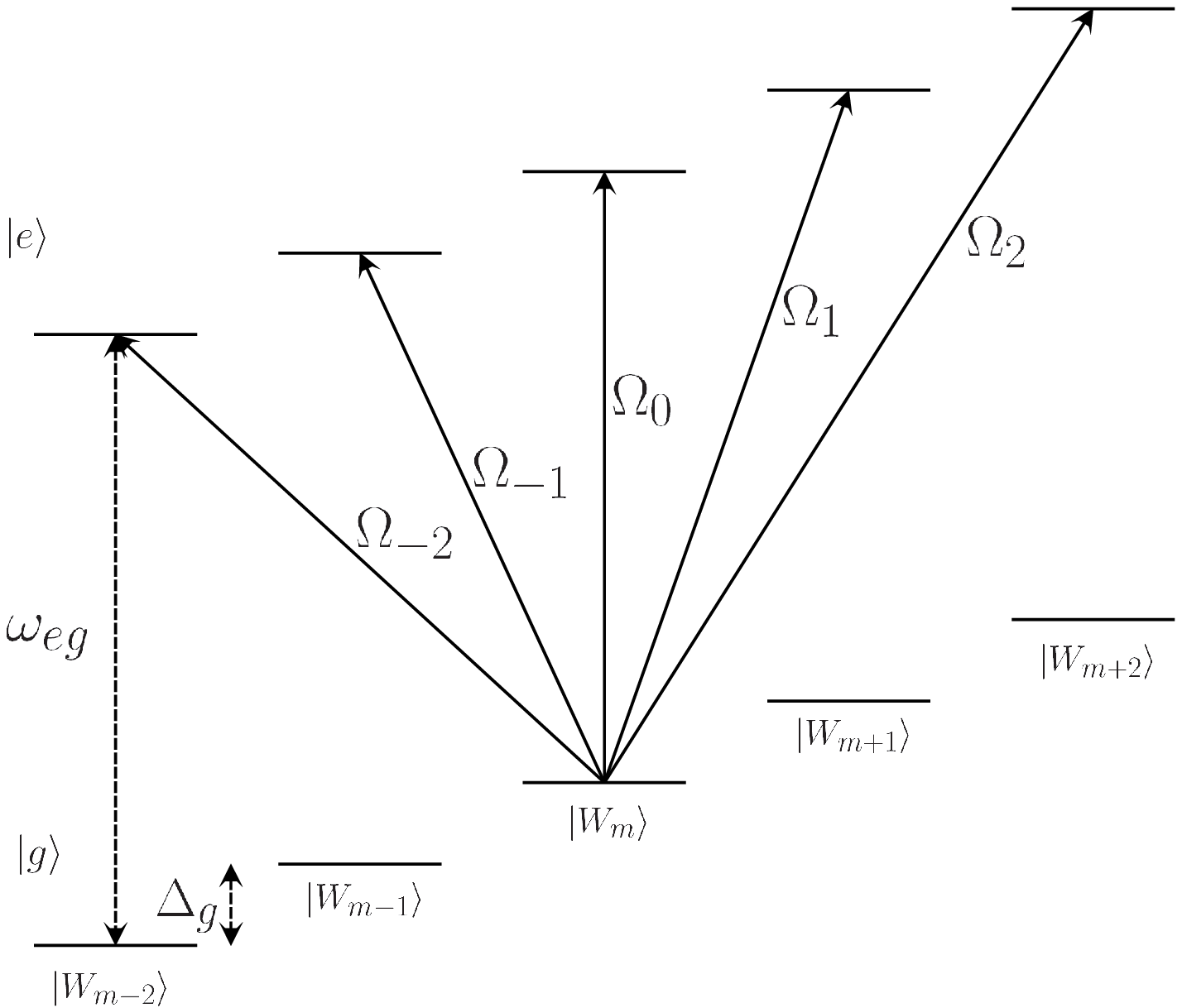}
\end{center}
  \caption{Wannier-Stark ladder of states and coupling between states by the probe laser.
  }\label{fig:wannier_stark_couplings}
\end{figure}

When the probe laser is switched on it couples $|W_m,g\rangle$ to $|W_{m'},e\rangle$ in either the same well ($m=m'$) or in neighboring wells by the translation in momentum space
$e^{ik_s\hat x}$, with the coupling strengths $\langle W_m|e^{ik_s\hat{x}}|W_{m'}\rangle$ (see figure \ref{fig:wannier_stark_couplings}).

\begin{figure}[b]
\begin{center}
  \includegraphics[width=5 cm]{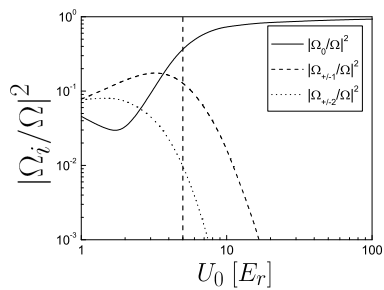}
\end{center}
  \caption{Relative coupling strength of the transition in the same well $|\Omega_0/\Omega|^2$ and transitions into the first four neighboring wells $|\Omega_{\pm 1}/\Omega|^2$ and $|\Omega_{\pm 2}/\Omega|^2$ as a function of the lattice depth $U_0$. The vertical dashed line corresponds to $U_0=5E_r$.}\label{fig:omegasgravity}
\end{figure}

Physically this process corresponds to tunneling through the periodic potential barriers induced by the probe laser, and depends strongly on the lattice depth $U_0$. Figure \ref{fig:omegasgravity} shows the relative coupling strengths for these processes as a function of $U_0$.

We note that at realistic lattice depths of a few $E_r$ the coupling strengths $\Omega_0$ (transition in the same well) and $\Omega_{\pm 1}$ (tunneling to neighboring wells) are of the same order. This implies that for a given probe laser intensity the two processes are equally likely, governed only by the frequency of the probe laser. For a probe laser linewidth that is significantly narrower than $\Delta_g$ (866\,Hz) atoms remain in the same well when the probe laser is on resonance ($\omega = \omega_{eg}$), climb one step up the WS-ladder when $\omega = \omega_{eg}+\Delta_g$ or one step down when $\omega = \omega_{eg}-\Delta_g$. These processes can be very efficient: in a numerical simulation with $U_0 = 5 E_r$ and $\Omega = 10$~Hz we obtain transition probabilities into neighboring wells that exceed 99.9\,\%. Consequently, a sequence of probe laser pulses on resonance or detuned by $\pm \Delta_g$, provide a powerful method of spatially separating and re-combining the atoms on the WS ladder. 

\section{Principle of the Experiment} \label{experiment}

\begin{figure}[b]
\begin{center}
\includegraphics[width=9 cm]{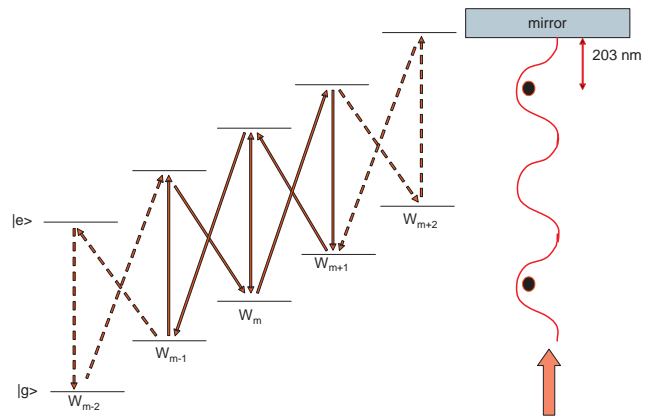}
\end{center}
\caption{Experimental principle: Vertical standing wave with atoms in the first and third well (right), and interferometer created by the sequence of pulses described in the text (left).}\label{fig:experiment}
\end{figure}

The experimental setup is sketched in figure \ref{fig:experiment} and is to some extent very similar to the setups used in optical lattice clocks \cite{Pierre} or for the measurement of Bloch oscillations \cite{Biraben,Tino}. The atoms are trapped in a vertical optical lattice with horizontal confinement provided by the Gaussian shape of the vertical trap laser or additional horizontal beams forming a 3D lattice. The probe laser is aligned and overlapped with the vertical trap laser. Typically, about $10^3$ to $10^4$ atoms per lattice site are trapped and cooled to a few $\mu$K.

The main experimental difference of the scheme proposed here with respect to existing setups is that atoms need to be trapped in lattice sites close to one of the mirrors of the vertical trap laser, and that provisions have to be made for selecting atoms in a single vertical well. The latter can be achieved by several means, all based on selective pumping of atoms from $|g\rangle$ to $|e\rangle$ and cleaning all remaining $|g\rangle$ states using a pusher beam. For example, a second vertical lattice superimposed onto the first
one but of different period would induce an additional light shift
dependent on the well index. Atoms in the maximally (or minimally)
light-shifted wells are then selectively shelved to $|e\rangle$
by a properly tuned probe pulse while atoms populating all the other
wells and remaining in $|g\rangle$ are cleared by the pusher beam. Depending on the period
ratio of both lattices and on the number of wells initially
populated, this leaves atoms in only one well or in a few wells
separated by a known number of sites. In addition since both
lattices are generated by reflection on the same surface the
index of the populated well(s) with respect to this surface can be
known unambiguously. Quite generally, this type of scheme can be used in the presence of any potential whose spatial gradient is sufficient to discriminate between neighboring wells using the probe laser.

Once atoms in a particular well are selected, they are transferred to the well of interest (at the distance at which the measurement is to take place) using a sequence of $\pi$ pulses of the probe laser detuned by $\pm\Delta_g$ to move up or down the WS-ladder as described in section \ref{control}. An interferometer is then created around that well using the following sequence (see figure \ref{fig:experiment}): We start with atoms in state $|W_m,g\rangle$. A first $\pi/2$ pulse on resonance creates a superposition of $|W_m,g\rangle$ and $|W_m,e\rangle$. Next, a $\pi$ pulse detuned by $+\Delta_g$ transfers atoms from $|W_m,g\rangle \rightarrow |W_{m+1},e\rangle$ and $|W_m,e\rangle \rightarrow |W_{m-1},g\rangle$ (those are the only transitions resonant with the probe laser detuned by $+\Delta_g$) leaving a superposition of spatially separated states in wells $m+1$ and $m-1$. After a time $T_1$ a "symmetrization" $\pi$ pulse on resonance switches internal states. A time $T_2$ later a $\pi$ pulse detuned by $-\Delta_g$ transfers atoms back ($|W_{m+1},g\rangle \rightarrow |W_{m},e\rangle$ and $|W_{m-1},e\rangle \rightarrow |W_{m},g\rangle$) with a final $\pi/2$ pulse on resonance recombining the atoms in the initial well $m$, where the internal state populations $P_e$ and $P_g$ are measured by fluorescence. The result is a "skewed butterfly" shaped interferometer (solid lines in figure \ref{fig:experiment}) of spatially separated paths with the final detection probabilities depending on the phase difference cumulated along the two paths. The different energies of the states along the two paths and the initial phases of the probe laser pulses lead to an overall phase difference

\begin{eqnarray}
	\Delta\phi=\frac{1}{\hbar} \ \left[\left(E_{m+1}-E_{m-1}\right)\left(T_1+T_2\right)\right] \nonumber \\
 +\left(\omega_{eg}^{(m+1)}T_1-\omega_{eg}^{(m-1)}T_2\right) \label{phasediff} \\
 -\phi_s^{(1)}+2(\phi_s^{(2)}-\phi_s^{(3)}+\phi_s^{(4)})-\phi_s^{(5)} \nonumber
\end{eqnarray}
where $E_{m}$ is the energy of state $|W_{m},g\rangle$, $\omega_{eg}^{(m)}$ is the separation between internal states in well $m$, and $\phi_s^{(i)}$ is the initial phase of the $i^{\rm \ th}$ pulse of the probe laser. Expressing $E_m$ as the known potential of the WS states plus an additional perturbation $U_m$ (QED, new interaction, stray e-m fields, etc...) the phase difference is

\begin{eqnarray}
	\Delta\phi=\frac{1}{\hbar} \ \left(m_ag\lambda_l+U_{m+1}-U_{m-1}\right)\left(T_1+T_2\right)\nonumber \\
 +\left(\omega_{eg}^{(m+1)}T_1-\omega_{eg}^{(m-1)}T_2\right)  \label{phasediff2} \\
 -\phi_s^{(1)}+2(\phi_s^{(2)}-\phi_s^{(3)}+\phi_s^{(4)})-\phi_s^{(5)}. \nonumber
\end{eqnarray}

As can be seen from equation (\ref{phasediff2}) the phase difference depends on the interactions times $T_i$ which leads to interference fringes as a function of $T_1+T_2$ at either output of the interferometer i.e. when measuring the internal state populations $P_e$ or $P_g$ after the last $\pi/2$ pulse. The signal of interest is in the first term of (\ref{phasediff2}) allowing the determination of $m_a/\hbar$ and $g$ (analogue of the measurements reported in \cite{Biraben,Roati,Biraben3,Tino}) or the measurement of any additional potential $U$ that varies over the size of the interferometer. The scheme can be modified to adapt to the size of the spatial variation of $U$ by adding additional $\pi$ pulses detuned by $\pm \Delta_g$ (dashed lines on figure \ref{fig:experiment}) leading to larger separations and limited only by the coherence time of the superposition and the efficiency of the transfer $W_m\rightarrow W_{m\pm 1}$.

Assuming state of the art measurement noise of atom interferometers we expect that $\Delta\phi$ can be determined with a precision of $\approx 10^{-4}$\,rad after $10^3$ to $10^4$\,s integration. For interaction times $T_1+T_2 \approx 0.1$\,s this corresponds to a measurement noise on $(m_ag\lambda_l+U_{m+1}-U_{m-1})/(2\pi\hbar)$ of about $1.6\ 10^{-4}$\,Hz.

The terms in the second and third lines of (\ref{phasediff2}) point to potential sources of uncertainty. In an ideal experiment the internal state separation $\omega_{eg}$ is spatially homogeneous and all pulses of the probe laser are phase coherent, leading to cancelation of those terms (when choosing $T_1=T_2$). In reality, perturbations of $\omega_{eg}$ are inhomogeneous (magnetic effects, light shifts, collisional shifts, etc...), which leads to imperfect cancelation of the two terms in the second line of (\ref{phasediff2}) and the phase noise of the probe laser contributes in the $\phi_s^{(i)}$ terms. We discuss these and other perturbing effects in the next section.

\section{Perturbations} \label{systematics}

For this section we will extend the discussion to a range of atoms (Li, Na, K, Rb, Cs, Mg, Ca, Sr, Yb, Hg), as the effect of some perturbations will significantly depend on the atom and isotope used. More precisely, we will consider two configuration for our interferometer using two level atoms in an optical lattice: the hyperfine splitting of the ground state (Li, Na, K, Rb, Cs) with counter propagating Raman pulses as the probe laser \cite{Biraben2,Biraben}, and optical transitions (Mg, Ca, Sr, Yb, Hg) \cite{Katori,Ye2,Pierre2,Barber}. Throughout the section we will investigate the potential perturbations with a $\approx 10^{-4}$\,rad goal of overall uncertainty on the measurement of $\Delta\phi$ in (\ref{phasediff2}) corresponding to perturbations of $\leq 10^{-4}$\,Hz in frequency.

\subsection{Phase coherence of the probe laser}
The third line in (\ref{phasediff2}) characterizes the effect of a phase incoherence of the different probe laser pulses. Assuming a cycle time of the experiment of 1\,s and data integration over $10^{4}$\,s, the $10^{-4}$\,rad goal implies a phase stability of the probe laser of $\approx 10^{-2}$\,rad over timescales of order 0.1 s, with no perturbations ($\geq 10^{-4}$\,rad) that are synchronous with the cycle time of the experiment.

When using Raman pulses with hyperfine transitions only the difference (GHz frequency) of the counter propagating Raman beams needs to be controlled at that level. This can be achieved when carefully controlling the microwave signal used to generate that difference. Also a second atomic cloud (see below) can be used if required.

For optical transitions the required phase coherence is difficult to achieve. The most promising approach is to use a second cloud of atoms in the same lattice (subjected to the same probe laser pulses) but far from the surface as a phase reference that allows to correct for the phase variations of the probe laser. A similar scheme when operating two atomic fountain clocks has lead to a reduction of the probe signal noise by a factor of 16 \cite{Bize2}. Even in that case a phase coherence of about 0.1\,rad is still required in order to unambiguously identify the central interference fringe in the measurement. Nonetheless, using this method a compensation of the phase fluctuations of the probe laser at the required level seems possible \cite{Bergquist} but remains a challenging task. 

\subsection{Light shifts}

Light shifts from the lattice laser will affect the measured phase difference in two ways. Firstly the induced modification of the transition frequency $\omega_{eg}$ leads to imperfect cancelation of the two terms in the second line of (\ref{phasediff2}) if the lattice laser intensity (and hence the induced light shift) varies in time between $T_1$ and $T_2$ and/or in space ($\omega_{eg}^{(m+1)}\neq\omega_{eg}^{(m-1)}$). Secondly a spatial variation of the intensity over the size of the interferometer will lead to a modification of the energy difference between the $|g\rangle$ states in the different wells, which modifies the $(U_{m+1}-U_{m-1})$ difference in (\ref{phasediff2}). The involved energy shifts are rather large when compared to the $10^{-4}$\,Hz goal, e.g. for Sr in a 5\,$E_r$ deep lattice the light shift of the $|g\rangle$ state is about 20\,kHz, which implies that the spatial variation of the lattice laser intensity needs to be controlled at the $10^{-8}$ level.

The hyperfine transition frequency is modified only by the {\it difference} of the light shifts of the two hyperfine states and is therefore about a factor $\omega_{hf}/\omega_{opt} \approx 10^{-5}$ smaller than the shift of each state \cite{note1}. Thus a relatively modest spatial (0.1 to 100\,$\mu$m) and temporal (0.1\,s) stability of the lattice laser at the  $10^{-4}$ level is sufficient. However, the induced shift of the $|g\rangle$ states is affected by the full light shift, and therefore the spatial variation of intensity over the lattice sites of interest needs to be controlled to about $10^{-8}$. For a quantitative estimate we will examine this effect for the experiment reported in \cite{Biraben}. In the picture of Bloch oscillations used in \cite{Biraben} a spatial variation of the trap laser intensity leads to a modification of the vertical wave vector $k_l$ ($k_B$ in \cite{Biraben}) as a function of vertical position, which modifies the momentum transferred by the Bloch oscillations. The results of \cite{Biraben} are consistent with $\delta k_l / k_l \leq 10^{-8}$. In our case, the separation of adjacent WS states $\Delta_g=m_ag\pi/(\hbar k_l)$ ($\approx 900$\,Hz), so the $10^{-8}$ uncertainty on $k_l$ leads to a contribution of about $10^{-5}$\,Hz, which is sufficient for our purposes. However, some caution is necessary when directly applying the results of \cite{Biraben} to our case: Firstly in \cite{Biraben} intensity variations are averaged over several wells as the atoms are delocalised. Secondly, and more importantly, controlling the intensity close to a reflecting surface is certainly more difficult than far from the surface (stray reflections, spurious modes, etc...). Nonetheless the results of \cite{Biraben} provide a good indication that light shifts due to spatial intensity fluctuations should be controllable at the required level.

For optical transitions the same conclusions apply concerning the shifts of the $|g\rangle$ states in the different wells. Concerning the spatial and temporal modification of $\omega_{eg}$ the problem is completely solved when the optical lattice is operated at the so-called "magic wavelength", which leads to insensitivity of $\omega_{eg}$ to the trap laser intensity to first order, and totally negligible higher order effects \cite{Pierre2}.

Finally, light shifts from wave front curvature and the Gouy phase have been studied in \cite{Biraben} and found to be consistent with $\delta k_l / k_l \leq 10^{-8}$, which again means that they should be compatible with the $10^{-4}$\,Hz uncertainty considered here.

\subsection{Collisional shift}

The interaction between atoms will affect the atomic energy levels as a function of the atomic density. As for the light shifts of the previous section, the effect will be two-fold, a modification of the transition frequency $\omega_{eg}$ (second line of (\ref{phasediff2})) and a modification of the ground state energy difference ($(U_{m+1}-U_{m-1})$ term in (\ref{phasediff2})).

For some bosonic isotopes considered here the dependence of $\omega_{eg}$ on density has been measured. It is reported in table \ref{tab:collisions} together with existing knowledge of the dominant ground state scattering lengths and the resulting frequency shift of the ground state using the relation $\Delta U = 4\pi\hbar^2 a \rho/m_a$ where $a$ is the scattering length and $\rho$ the atomic density.

\begin{table}
\caption{\label{tab:collisions}Dominant scattering lengths and collisional shifts for bosonic isotopes. $\Delta U$ and $\delta\omega_{eg}$ are given for an atomic density of $10^{12}$ atoms cm$^{-3}$.}
\begin{ruledtabular}
\begin{tabular}{cccccc}
Atom & $a$ & $\Delta U$ & $\frac{\delta\omega_{eg}}{\rho}$ & $\delta\omega_{eg}$ & Refs. \\
 & [nm] & [Hz] & [Hz cm$^3$] & [Hz] &  \\
\hline
$^{87}$Rb & 5 & 7 & $3 \times 10^{-13}$ & $0.3$ &  \cite{Marte,Fertig,Sortais}\\
$^{133}$Cs & 127 & 122 & $2\times 10^{-11}$ & 20 & \cite{Leo1,Leo2} \\
$^{40}$Ca & 2.6 - 16 & $\approx$29 & $\leq 7\times 10^{-11}$ & $\leq 70$ & \cite{Degenhardt,Sterr} \\
$^{7}$Li & 1.7 & 30 & - & - & \cite{Abraham} \\
$^{23}$Na & 1.0 & 5.4 & - & - & \cite{Samuelis} \\
$^{39}$K & 7.4 & 24 & - & - & \cite{Burke,Williams} \\
$^{24}$Mg & 1.4 & 7.3 & - & - & \cite{Tiesinga} \\
$^{86}$Sr & 32 - 120 & $\approx$100 & - & - & \cite{Killian} \\
$^{88}$Sr & $< 0.7$ & $< 1$ & - & - & \cite{Killian} \\
$^{174}$Yb & 1 - 3 & $\approx$1.5 & - & - & \cite{Takasu} \\
\end{tabular}
\end{ruledtabular}
\end{table}

Using techniques based on adiabatic transfer \cite{Bize} the collisional shift $\delta\omega_{eg}$ can be controlled to, at present, about $2\times 10^{-3}$, so table \ref{tab:collisions} indicates that none of the considered bosonic isotopes satisfy the $10^{-4}$\,Hz uncertainty considered here \cite{note2}. Furthermore, the values of $\Delta U$ (of the order of 1 Hz in the best cases) indicate that the density difference between the two wells of the superposition needs to be controlled to $\leq 10^{-4}$ which seems difficult to achieve. 

However, for fermionic isotopes collisional effects are orders of magnitude smaller (when spin polarized) due to the Pauli exclusion principle \cite{Gupta}. Fermionic isotopes exist for many of the considered atoms ($^{6}$Li, $^{40}$K, $^{25}$Mg, $^{43}$Ca $^{87}$Sr, $^{171,173}$Yb, $^{199,201}$Hg) which provides the possibility of carrying out the experiment using fermions.

\subsection{Knowledge of $m_ag\lambda_l/\hbar$}

A $10^{-4}$\,rad measurement of $\Delta\phi$ requires the knowledge of the first term in (\ref{phasediff2}) at or below the $10^{-7}$ level in relative uncertainty. 

For $\hbar/m_a$ this is already achieved for several atoms. For example $\hbar/m_{Rb}$ is determined with a relative uncertainty of $1.4\times 10^{-8}$ \cite{Biraben}, and $\hbar/m_{Cs}$ with $1.6\times 10^{-8}$ \cite{Wicht}. Even if this were not the case, the experiment itself could be used to first measure $\hbar/m_a$ at the required uncertainty ($\leq 10^{-7}$) using an atomic cloud far from the surface.

The trap laser wavelength $\lambda_l$ can be known to much better than $10^{-7}$ even with modest stabilization and measurement techniques.

The absolute value of local gravity $g$ needs to be determined at the location of the atoms, including any vibrations and other perturbations (tides, gravity gradient, etc...). Absolute gravimeters routinely reach $< 10^{-8}\,g$ uncertainty, vibration isolations that achieve residual fluctuations of $\leq 10^{-6}\,g/\sqrt{{\rm Hz}}$ at frequencies around 10\,Hz are commonly used in gravimetry, and tidal models are largely sufficient to correct tidal effects at that level.

\subsection{Stray electric and magnetic fields}

One of the most significant error sources in the measurement reported in \cite{Cornell} are stray electric and magnetic fields originating from contaminations of the surface. The resulting normalized perturbation of the measured center of mass oscillation of the BEC is of the order $10^{-5}$ (see table I in \cite{Cornell}), corresponding to an uncertainty of around $10^{-2}$\,Hz on the measurement of a potential with $r^{-4}$ dependence (e.g. Casimir-Polder potential, $U_{CP}$) at a distance of $\approx 7.5\,\mu$m from the surface. The surface charges responsible for the effect are either spurious charges (especially when using an insulating surface) or dipoles on the surface created by adsorbed atoms (especially for conducting and semiconducting surfaces) as extensively studied in \cite{Cornell_2}.

It is difficult to realistically estimate the corresponding effect for our case because of the different experimental conditions that significantly affect the stray fields (number of "stray" atoms, necessity for high vacuum, used surfaces and atoms, etc...). However, we believe that most of these differences lead to less sensitivity to stray fields (see below) with good hope of controlling them to the required level, especially at relatively large distances from the surface (5 to 10\,$\mu$m).

As noted in \cite{Cornell} conducting or semiconducting surfaces would be preferable to insulators largely because they are less susceptible to electric fields caused by spurious surface charges. However, they were not used because of Rb atoms sticking to the surface and creating electric dipoles \cite{Cornell_2}. In our case this is likely to be less of a problem for several reasons: The residual pressure in the proposed experiment can be about 100 times higher than required for typical BEC experiments, which means that it is possible to keep the surface clean by heating it to temperatures that are incompatible with the ultra high vacuum required for BECs. The total number of atoms required to produce a BEC is about $10^3$ times larger than for the case of $\mu$K atoms in an optical lattice (for the same number of trapped atoms), which implies that less atoms will adsorb onto the surface in our case. Also, other means for protecting the surface against adsorption could be used, e.g. a blue-detuned evanescent wave like in \cite{Arnaud}, and finally for non-alkaline atoms the strength of the dipoles created by adsorption might be significantly less than for Rb studied in \cite{Cornell_2}.

Concerning magnetic fields, in \cite{Cornell} it was not possible to apply an external magnetic field in order to detect their effect (analogously to the method used for the $E$-field) because of the magnetic trap used for the BEC. This is not the case for the purely optical trap proposed here, so characterization of magnetic effects is likely to be more precise than in \cite{Cornell}, especially for isotopes with non-zero nuclear angular momentum (all fermionic isotopes and alkaline bosons) as the different $m_F$ states can then be used to measure the magnetic fields "in situ".

\subsection{Knowledge of the atom-surface separation}

The uncertainty of any measurement of atom-surface interactions depends crucially on the precise determination of the distance between the atom and the surface. For example, a \% measurement of the Casimir-Polder potential ($U_{CP}$) requires a $10^{-3}$ uncertainty on $r$ (due to the $r^{-4}$ dependence of $U_{CP}$), which at the short distances involved is no trivial task. In the most recent experiment \cite{Cornell} at relatively large distances ($6-10\,\mu$m) the uncertainties on $r$ are around 0.2\,$\mu$m (see figure 2 in \cite{Cornell}) which implies a limit of about 10\,\% on the measurement of $U_{CP}$.

This is where we see one of the main advantages of the method proposed here. The precise knowledge of the external state of the atoms localized in WS states, and of the position of the optical lattice with respect to the surface allows an accurate determination of the position of the WS states (and hence of the atoms) with respect to the reflecting surface.  

The uncertainty of the trap laser wavelength contributes negligibly to $\delta r$ (below a picometre), but the characterization of its wave-fronts may have a larger effect. Over the small extension of the trapped atoms ($\approx 100 \mu$m) it should be possible to control wave front curvatures to about $10^{-3}\,\lambda_l$ leading to $\delta r \approx 0.8$\,nm. However, interference between the trap laser and stray reflections due to surface roughness may play a non-negligible role \cite{Henkelthese,HenkelPRA}.

Other effects could come from the surface itself. Surface roughness contributes to $\approx 0.5$\,nm \cite{Cornell} and is the dominant effect. The material properties of the surface need to be taken into account when calculating $U_{CP}$ \cite{Astrid2000,Astrid2006} using the frequency dependent reflectivity of the surface. For metallic surfaces this should be possible at the required $10^{-3}$ to $10^{-2}$ level when measuring the optical properties of the used surface \cite{Astrid2006}, especially at distances larger than the plasma wavelength, but it is likely to be more difficult for coated dielectrics.

Thus, we believe that our experimental method will allow the control of the atom-surface separation to $\approx$\,1\,nm, limited by the surface roughness and the wave-front imperfections of the lattice.

In summary, the effect of the perturbations discussed above will depend significantly on the choices made for the atoms and the surface. Our estimate for the collisional shift seems to exclude most bosonic isotopes but is not expected to affect fermionic isotopes, and the likely effect of stray magnetic and electric fields as well as the required knowledge of the reflectivity seems to favor a conducting or semiconducting surface. Under those conditions a control of all perturbations of $\Delta\phi$ in (\ref{phasediff2}) at the $10^{-4}$ rad level seems possible but remains challenging, especially concerning the phase coherence of the probe laser and the control of stray electromagnetic fields. We therefore take this value as the basis for the discussion of the scientific interest of the experiment in the next two sections.

\section{QED Interaction} \label{QED}

After its prediction in 1948 \cite{Casimir}, the Casimir force
has been observed in a number of "historic" experiments which confirmed its existence and main properties
\cite{Sparnaay,Milonni,Mostepanenko,Lamoreaux}. More recent measurements \cite{Bordag} with largely improved accuracy have allowed for comparison between measured values of the force and theoretical predictions at the few \% level. This comparison is interesting because the Casimir force is the most accessible effect of vacuum fluctuations in the macroscopic world. As the existence of vacuum energy raises difficulties at the interface
between the theories of quantum and gravitational phenomena, it is worth testing this effect with the greatest
care and highest accuracy \cite{Reynaud,Genet}.

Shortly after the prediction of the Casimir force between two parallel plates, Casimir and Polder \cite{CP} predicted the analogous attractive force between an atom and a macroscopic plane surface. The corresponding potential is known as the Casimir-Polder potential $U_{CP}$ and is described to first approximation (zero temperature, perfect conducting surface, only static part of the atomic polarizability, etc...) by

\begin{equation}
	U_{CP} = \frac{3\hbar c \alpha_0}{8\pi r^4} \label{UCP}
\end{equation}
where $c$ is the speed of light in vacuum, $\alpha_0$ is the static polarizability of the atom, and $r$ is the distance between the atom and the surface. More generally, the Casimir-Polder potential is the retarded part of the total QED interaction $U_{QED}$ between the atom and the surface. The non-retarded part, which is dominant at short distances, is known as the "Van der Waals" potential $U_{VdW}$ and has a $r^{-3}$ dependence. Also, at non-zero temperatures (and larger distances), the overall QED potential becomes dominated by a temperature dependent term again with $r^{-3}$ dependence of the form (to first approximation)

\begin{equation}
	U_{T} = \frac{k_B T \alpha_0}{4 r^3} \label{T}
\end{equation}
where $k_B$ is the Boltzmann constant and $T$ the temperature.

This leads to an interesting phenomenological behavior with two distance dependent crossover points from $r^{-3}$ dependence to $r^{-4}$ and back. At 300\,K the two crossover points are situated at a few tenths of a micron and a few microns respectively (depending on the atom and surface characteristics). Experiments that measure the overall QED interaction use cold atoms \cite{Arnaud,Shimizu,Oberst} or BECs \cite{Lin,Pasquini,Cornell}. The experiment by Sukenik \cite{Sukenik} was the first to clearly observe the crossover between $U_{VdW}$ and $U_{CP}$, whereas the experiment by Harber et al. \cite{Cornell} has concentrated on the second crossover at larger distances (from $U_{CP}$ to $U_T$), without clear evidence so far. Typically, experiments measuring the atom-wall QED interaction have an overall relative uncertainty at or above 10\,\% (e.g. the overall uncertainty of \cite{Arnaud} is estimated to about 30\,\%).

Our proposed experiment (section \ref{experiment}) measures the potential difference between two wells separated by a distance of $n\lambda_l$ ($2n$ wells) with $n=1,2,3,...$ depending on the number of $\pi$ pulses applied for the separation of the atoms. For wells that are close to the surface that potential difference is dominated by $U_{QED}$. 

\begin{table}
\caption{\label{tab:UCP} QED potential for Sr in different wells of the optical lattice at 813\,nm. $U$ is evaluated approximately using (\ref{UCP}) for wells 1 to 5 and (\ref{T}) for well 26.}
\begin{ruledtabular}
\begin{tabular}{cccccccc}
Well no. & 1 & 2 & 3 & 4 & 5 &   & 26 \\
\vspace{1mm}
$r$\,/nm & 203 & 610 & 1016 & 1423 & 1829 &  & 10366 \\
\vspace{1mm}
$\frac{U}{2\pi\hbar}$\,/Hz & $(10^{5})$ & 1200 & 160 & 41 & 15 &   & 0.04\\
\end{tabular}
\end{ruledtabular}
\end{table}

Table \ref{tab:UCP} shows the QED potential evaluated using approximations (\ref{UCP}) and (\ref{T}) for Sr atoms in a lattice with $\lambda_l=813$\,nm. We first note that the proposed depth of the potential wells ($U_0=5E_r$) corresponds to about 20\,kHz, i.e. less than $U_{QED}$ in the first well, which means that atoms cannot be trapped in that well. Even in the next two wells or so, $U_{QED}$ is far from being a "small" perturbation of the trapping potential and will significantly modify the WS states and transitions probabilities calculated in \cite{LW} (see sections \ref{WS}, \ref{control}) in the absence of any additional potential. Nonetheless, the principle of the experiment remains valid but it requires a full quantum calculation (including $U_{QED}$) to obtain the correct WS states and corresponding energy differences, transition probabilities, etc.... As a result, given the non-negligible shift of transition frequencies to neighboring wells ($\approx$\,kHz for the 2nd well), some experimental complications may be necessary when exploring wells very close to the surface, for example, using simultaneously two slightly detuned probe lasers. 

The measurement of $U_{QED}$ using the proposed experiment is now straightforward. For a measurement at short distance, we prepare the atoms in the 4th well for example, and then use the sequence of pulses described in section \ref{experiment} to create a superposition between wells 3 and 5. Our projected uncertainty of $10^{-4}$\,Hz then leads to a measurement of $U_{CP}$ at a relative uncertainty of about $10^{-6}$, with potential for another order of magnitude improvement when starting in well no. 3 (superposition between 2 and 4). However, at such short distances the main limitation will come from the uncertainty on the determination of the atom-surface separation ($\delta r \leq 1$\,nm see section \ref{systematics}) which for atoms in well 3 leads to a contribution of about $\delta U/U = 4\delta r/r \approx 0.04$ and worse when closer to the surface. At large distances (e.g. for a superposition between the 26th and 40th well) the effect of $U_T$ is about 0.03\,Hz, which implies a measurement at $\delta U/U < 0.01$ now limited by the $10^{-4}$\,Hz uncertainty in the measurement of $U$ rather than the uncertainty in $r$. The optimum is $\delta U/U \leq 10^{-3}$ situated at intermediate distances of about 5 microns from the surface, close to the crossover between (\ref{UCP}) and (\ref{T}) at about 3.6 microns.

Thus, a measurement at optimum distance corresponds to two orders of magnitude improvement on the experimental verification of $U_{QED}$ between an atom and a macroscopic surface, down to the $10^{-3}$ level in relative uncertainty. This is of interest, given that theoretical predictions for real conditions have typical uncertainties of some percent \cite{Astrid2006,Babb} (depending on how well the material properties of the mirrors are known \cite{Astrid2006}), and that recent theoretical work \cite{Antezza} suggests new phenomenological behavior of $U_{QED}$ out of thermal equilibrium at intermediate and large distances (few microns and above). Finally, we point out that our method also allows a complementary measurement using direct spectroscopy of $\omega_{eg}$ as a function of the distance from the surface (clock operation with atoms prepared in a given well), which allows the exploration of $U_{QED}$ as a function of distance and internal state of the atom.

\section{Search for New Interactions} \label{YukS}

The search for deviations from Newton's law of gravitation has been a recurrent issue for the
last three decades. Initially motivated by the possibility of deviations from standard gravity
due to new forces with couplings of the order of that of gravity \cite{Fischbach}, the search has been more
recently encouraged by unification models which predict the existence of forces up to $10^5$
times stronger than gravity in the 1\,$\mu$m and 100\,$\mu$m range \cite{Long,Adelberger}. Even if its results have
not met initial hopes for the observation of a "fifth force", the search has greatly improved
understanding of gravitation, generated an impressive body of new knowledge, and
narrowed the remaining open windows for new fundamental forces.

\begin{figure}[b]
\begin{center}
\includegraphics[width=8 cm]{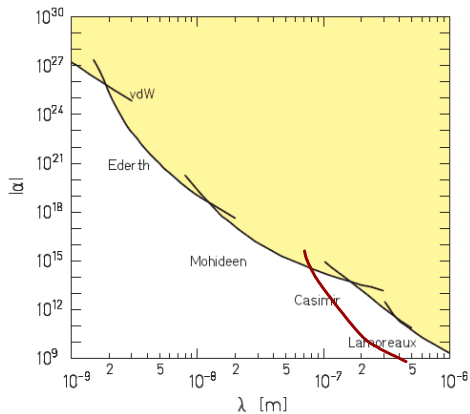}
\end{center}
\caption{Limits \cite{comm_limits} on $U_{Yuk}$ at very short ranges (adapted from \cite{Adelberger}). The solid red line indicates estimated limits from the present proposal when using a differential measurement between isotopes.}\label{limits_short}
\end{figure}

\begin{figure}[b]
\begin{center}
\includegraphics[width=8 cm]{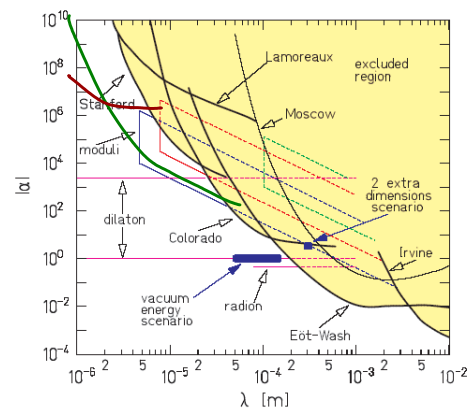}
\end{center}
\caption{Limits \cite{comm_limits} on $U_{Yuk}$ at short ranges (adapted from \cite{Adelberger}). The solid red line is as for figure \ref{limits_short}. The solid green line indicates expected limits from the present proposal with atoms at $\approx 10 \mu$m from the surface.}\label{limits_long}
\end{figure}

The hypothetical additional gravitational potential $U_{Yuk}$ is often expressed in the form of a Yukawa potential and parametrised by a dimensionless coupling strength $\alpha$ relative to the Newtonian gravitational potential, and a range $\lambda$

\begin{equation}
	U_{Yuk} = U_N (\alpha e^{-r/\lambda}) \label{Yuk}
\end{equation}
where $U_N$ is the standard Newtonian potential. Experiments set limits in the $(\lambda,\alpha)$ plane, each experiment  providing best limits in the region of $\lambda$ corresponding to the typical distances between the masses used. For large $\lambda$ ($10^5$ to $10^{15}$\,m) limits are provided by artificial satellite, lunar, and solar system observations, whereas ranges of 1\,m and below are the domain of laboratory experiments. The knowledge of $U_{Yuk}$ deteriorates rapidly (limits on $\alpha$ increase) for very large ($>10^{15}$\,m) and very small ($<10^{-3}$\,m) ranges, leaving "open windows", of which the small $\lambda$ one is the focus of this work. Figures \ref{limits_short} and \ref{limits_long} (adapted from \cite{Adelberger}) show the present limits at very short ($10^{-9}$ to $10^{-6}$\,m) and short range ($10^{-6}$ to $10^{-2}$\,m). At a range of about 2 microns best limits can only exclude an additional potential larger than about $10^{10}$ times gravity (figure \ref{limits_long}, \cite{comm_limits}) with much worse results at shorter distances. Figure \ref{limits_long} also shows some theoretical predictions, which seem to indicate that some interesting ranges to explore are between a micron and a millimeter.

The interest of the experiment proposed here in measuring such short range potentials is two-fold: the large region of distances that can be explored by preparing the atoms in different wells of the optical lattice, and the comparatively low uncertainty of the measurement itself (measuring directly the potential rather than the force). For example, with a measurement in a superposition between the 3rd and 7th wells (Sr in a lattice with $\lambda_l=813$\,nm, Au surface, $T_1+T_2=0.1$\,s) and a measurement uncertainty of $10^{-4}$\,rad in (\ref{phasediff2}) we obtain a limit of $\alpha \leq 10^{4.8}$ at $\lambda=2$\,$\mu$m, around 5 to 6 orders of magnitude better than previous limits (figure \ref{limits_long}, \cite{comm_limits}). At larger distances (26th and 50th well) the limit is $\alpha \leq 10^{3.6}$ at $\lambda=10$\,$\mu$m, about a two orders of magnitude improvement.

Contrary to the measurement of $U_{QED}$ (section \ref{QED}) the uncertainty in the knowledge of the atom-surface separation only plays a minor role as, unless we discover a non-zero potential, we only set an upper limit on $U_{Yuk}$ i.e. we require that $\delta U_{Yuk}/U_{Yuk} \leq 1$, which implies that $\delta r/\lambda\leq 1$, largely within our uncertainty (see section \ref{systematics}).

In fact, the most serious issue when trying to measure $U_{Yuk}$ is the perturbation from $U_{QED}$, especially at short distances. Table \ref{tab:UCP} shows that for measurements in the 3rd well one needs to correct and/or cancel the effect of $U_{QED}$ at the $10^{-6}$ level for the $10^{-4}$\,Hz uncertainty that we aim at, and even when relatively far from the surface (26th well) a correction at the \% level is still required. We believe that the most reasonable approach to that limitation is a two stage experiment, starting with an experiment at relatively large distance ($\geq 10\,\mu$m) with the setup described above, and exploring shorter separations in a second stage where experimental precautions are taken to cancel and control $U_{QED}$ at the required level.

For the first stage, $U_{QED}$ needs to be calculated and corrected at the \% level to reach the required $10^{-4}$\,Hz uncertainty (c.f. table \ref{tab:UCP}) which is within the reach of present theory \cite{Astrid2006,Babb}, especially at such relatively large distances where the material properties of the mirrors are not crucial any more. So the experiment will be limited directly by the overall uncertainty on $\Delta \phi$ ($\approx 10^{-4}$\,rad) in (\ref{phasediff2}) leading to the curve at large distances in figure \ref{limits_long}, and an overall improvement by about 3 to 4 orders of magnitude on present limits.

The second stage requires one or several experimental "tricks" to reduce the effect of $U_{QED}$. For example, one could use a mirror with a narrow band reflectivity (around $\lambda_l$), or on the contrary a transparent (especially at the main atomic transition frequencies) source mass placed between the mirror and the atom whose distance to the mirror (and hence to the atom) is measured using an additional interferometer. Both methods could at most reduce the effect of $U_{QED}$ by about two orders of magnitude (the order of residual reflectivity or transparency of available materials). The more promising approach (that could also be combined with one of the previous two) is to carry out a differential measurement between two isotopes of the same atom (e.g. $^{85,87}$Rb or $^{171,173}$Yb). The difference in $U_{QED}$ between two isotopes is determined by the difference in polarizability, which is expected to be of the same order as the isotope shift of the main atomic transitions (typically around $10^{-6}$ in relative value \cite{Courtillot}), so a differential measurement should allow the cancelation of $U_{QED}$ at about that level, consistent with the required uncertainty for a measurement in the third well or further. The disadvantage of such a measurement is that only the differential effect of $U_{Yuk}$ is observed which is a factor $\Delta m_{at}/m_{at} \approx 0.01$ (with $\Delta m_{at}$ the isotopic mass difference) smaller than the full effect. Taking that into account, we obtain the projected limits at very short and intermediate distances shown in figures \ref{limits_short} and \ref{limits_long}, with an improvement of about 3 orders of magnitude on best previous results.   

\section{Conclusion} \label{conclusion}

We have proposed a novel experiment to measure the interaction between an atom and a macroscopic surface at short distances based on the existing technology of optical lattice clocks, with atoms trapped in lattice sites close to one of the reflecting surfaces. Our detailed study of perturbing effects (section \ref{systematics}) indicates that carrying out the experiment with present day technology should allow improvements of $\geq 2$ orders of magnitude on the knowledge of the atom-surface QED interaction (section \ref{QED}) and of up to four orders of magnitude on the limits of new short range interactions related to gravity (section \ref{YukS}).

The fundamental advantages of the described experiment are the possibility to determine the atom-surface separation with good accuracy ($\approx 1$\,nm), and the large range over which experimental parameters can be varied. For example, by placing the atoms in different wells distances from 600 nm to several tens of microns can be explored, and using different isotopes and internal states allows the study of the dependence of $U_{QED}$ and $U_{Yuk}$ on those parameters. The distance ranges of best sensitivity (between 1 and 10 microns) are of particular interest as they correspond to the transition between the Casimir-Polder and thermal regime in $U_{QED}$ (see section \ref{QED}) and to the region where several theoretical predictions of new interactions can be found (see figure \ref{limits_long}). We also note that the possibility of varying many experimental "knobs" (trap laser intensity and wavelength, atomic density, temperature, etc...) provides good handles to study and characterize many of the systematic effects discussed in section \ref{systematics}.

Of the studied atoms the most promising candidates are fermionic isotopes ($^{6}$Li, $^{40}$K, $^{25}$Mg, $^{43}$Ca $^{87}$Sr, $^{171,173}$Yb, $^{199,201}$Hg) because of the expected low energy shifts due to collisions when spin polarized. Of those only $^{6}$Li and $^{40}$K allow the use of a hyperfine transition which significantly relaxes the constraints on the coherence of the probe laser when using Raman pulses. However, the comparatively small mass of $^{6}$Li implies a corresponding reduction of the signal in the search for new interactions related to gravity, and the low natural abundance of $^{40}$K (0.01 \%) makes experiments more difficult. For the other fermions (Mg, Sr, Yb, Hg) the possibility of operating the optical lattice at the "magic" wavelength allows cancelation of the light shift of $\omega_{eg}$, but the optical clock transition imposes relatively stringent constraints on the coherence of the probe laser. Finally, the presence of two fermionic isotopes (Yb, Hg) allows the cancelation of $U_{QED}$ between isotopes which makes them good candidates for the search for new interactions at very short distances (see section \ref{YukS}). Concerning the material to be used for the reflecting surface, the likely effect of stray magnetic and electric fields as well as the required knowledge of the reflectivity seems to favor a conducting or semiconducting surface, but some care has to be taken to avoid adsorption of atoms on the surface that would form dipoles and corresponding stray fields (see section \ref{systematics}).

In summary, under the above conditions the experiment is feasible and, with state of the art technology, should lead to significant improvements in the knowledge of short range interactions between an atom and a macroscopic surface. Furthermore, even in a preliminary configuration where performances would be about two orders of magnitude lower than their ultimate limits (e.g. using Rb with a 10 to 100 fold loss in sensitivity due to collisions), the experiment would still be of interest when compared to previous measurements (see section \ref{QED}, \ref{YukS}) whilst using an experimental approach that is  fundamentally different from previous mechanical measurements, which is of essence in experimental investigations of fundamental physics.

\bibliographystyle{prsty}

\begin{thebibliography}{10}
\bibitem{Dimopoulos} S. Dimopoulos, A.A. Geraci, Phys.Rev.{\bf D68}, 124021, (2003).
\bibitem{Fischbach} E. Fischbach, C.L. Talmadge, {\it The Search for Non-Newtonian Gravity}, Springer, New York, (1999).
\bibitem{FischbachPRD} E. Fischbach, D.E. Krause, V.M. Mostepanenko, M. Novello, Phys.Rev.{\bf D64}, 075010, (2001).
\bibitem{Hoyle} C.D. Hoyle, et al., Phys.Rev.Lett. {\bf 68}, 1418, (2001).
\bibitem{Sukenik} C.I. Sukenik, et al., Phys.Rev.Lett. {\bf 70}, 560, (1993).
\bibitem{Arnaud} A. Landragin, Phys.Rev.Lett. {\bf 77}, 1464, (1996).
\bibitem{Shimizu} Phys. Rev. Lett. {\bf 86}, 987, (2001).
\bibitem{Druzhinina} V. Druzhinina, M. DeKieviet, Phys. Rev. Lett. {\bf 91}, 193202, (2003).
\bibitem{Lin} Y. J. Lin, I. Teper, C. Chin, V. Vuletic, Phys. Rev. Lett. {\bf 92}, 050404, (2004).
\bibitem{Pasquini} T. A. Pasquini, et al., Phys. Rev. Lett. {\bf 93}, 223201, (2004).
\bibitem{Oberst} H. Oberst, Y. Tashiro, K. Shimizu, F. Shimizu, Phys. Rev. {\bf A71}, 052901, (2005).
\bibitem{Cornell} D.M. Harber, J.M. Obrecht, J.M. McGuirk, E.A. Cornell, Phys.Rev. {\bf A72}, 033610, (2005).
\bibitem{Inguscio} I. Carusotto, et al., Phys.Rev.Lett. {\bf 95}, 093202, (2005).
\bibitem{Tino} G. Ferrari, N. Poli, F. Sorrentino, G.M. Tino, arXiv:physics/0605018.
\bibitem{Biraben2} R. Battesti et al., Phys. Rev. Lett. {\bf 92}, 253001, (2004).
\bibitem{Biraben} P. Clad\'e, et al., Phys. Rev. Lett. {\bf 96}, 033001, (2006).
\bibitem{Katori} M. Takamoto, F.-L. Hong, R. Higashi, H. Katori, Nature (London) {\bf 435}, 321, (2005).
\bibitem{Ye2} A.D. Ludlow, et al., Phys. Rev. Lett. {\bf 96}, 033003, (2006).
\bibitem{Pierre2} A. Brusch, et al., Phys. Rev. Lett. {\bf 96}, 103003, (2006).
\bibitem{Barber} Z.W. Barber et al., Phys. Rev. Lett. {\bf 96}, 083002, (2006).
\bibitem{Pierre} R. Le Targat, et al., arXiv:physics/0605200.
\bibitem{LW} P. Lemonde, P. Wolf, Phys. Rev. {\bf A72}, 033409, (2005).
\bibitem{Nenciu} G. Nenciu, Rev. Mod. Phys. {\bf 63}, 91, (1991).
\bibitem{Roati} G. Roati, et al., Phys. Rev. Lett. {\bf 92}, 230402, (2004).
\bibitem{Biraben3} P. Clad\'e, et al., Europhysics Letters {\bf 71}, 730, (2005).
\bibitem{Bize2} S. Bize, et al., IEEE Trans. UFFC {\bf 47}, 1253, (2000).
\bibitem{Bergquist} Young B.C. et al., Phys. Rev. Lett. {\bf 82}, 3799, (1999).
\bibitem{note1} For example, in Cs the shift of $\omega_{eg}$ is measured to be $4.6 \times 10^{-5}$ of the shift of $|g\rangle$ as obtained from the measured static polarizability \cite{Simon,Amini}.
\bibitem{Simon} E. Simon, P. Laurent, and A. Clairon, Phys. Rev. {\bf A57}, 436, (1998).
\bibitem{Amini} J.M. Amini, H. Gould, Phys. Rev. Lett. {\bf 91}, 153001, (2003).
\bibitem{Marte} A. Marte, et al., Phys. Rev. Lett. {\bf 89}, 283202, (2002).
\bibitem{Fertig} C. Fertig, K. Gibble, Phys. Rev. Lett. {\bf 85}, 1622, (2000).
\bibitem{Sortais} Y. Sortais, et al., Phys. Rev. Lett. {\bf 85}, 3117, (2000).
\bibitem{Leo1} P.J. Leo, C.J. Williams, P.S. Julienne, Phys. Rev. Lett. {\bf 85}, 2721, (2000).
\bibitem{Leo2} P.J. Leo, P.S. Julienne, F.H. Mies, C.J. Williams, Phys. Rev. Lett. {\bf 86}, 3743, (2001).
\bibitem{Degenhardt} C. Degenhardt, et al., Phys. Rev. {\bf A67}, 043408, (2003).
\bibitem{Sterr} U. Sterr, et al., Comptes Rendus Physique {\bf 5}, 845, (2004).
\bibitem{Abraham} E.R.I. Abraham, et al., Phys. Rev. {\bf A55}, 3299(R), (1997).
\bibitem{Samuelis} C. Samuelis, et al., Phys. Rev. {\bf A63}, 012710, (2000).
\bibitem{Burke} J.P. Burke, et al., Phys. Rev. {\bf A60}, 4417, (1999).
\bibitem{Williams} C.J. Williams, et al., Phys. Rev. {\bf A60}, 4427, (1999).
\bibitem{Tiesinga} E. Tiesinga, S. Kotochigova, P.S. Julienne, Phys. Rev. {\bf A65}, 042722, (2002).
\bibitem{Killian} P.G. Mickelson, et al., Phys. Rev. Lett. {\bf 95}, 223002, (2005).
\bibitem{Takasu} Y. Takasu, et al., Phys. Rev. Lett. {\bf 91}, 040404, (2003).
\bibitem{Bize} S. Bize, et al., J. Phys. B: At. Mol. Opt. Phys. {\bf 38}, S449, (2005).
\bibitem{note2} Note however, that the measurements of $\delta\omega_{eg}$ for $^{40}$Ca and the determination of $a$ for $^{88}$Sr are compatible with 0 within the uncertainties.
\bibitem{Gupta} S. Gupta et al., Science, {\bf 300}, 1723, (2003).
\bibitem{Wicht} A. Wicht, et al., Phys. Scr. {\bf T102}, 82, (2002).
\bibitem{Cornell_2} J.M. McGuirk, D.M. Harber, J.M. Obrecht, E.A. Cornell, Phys.Rev. {\bf A69}, 062905, (2004).
\bibitem{Henkelthese} C. Henkel, Ph.D. thesis, (1996), available on http//tel.ccsd.cnrs.fr/tel-00006757.
\bibitem{HenkelPRA} C. Henkel, et al., Phys. Rev. {\bf A55}, 1160, (1997).
\bibitem{Astrid2000} A. Lambrecht, S. Reynaud, Eur. Phys. J. {\bf D8}, 309, (2000).
\bibitem{Astrid2006} I. Pirozhenko, A. Lambrecht, V.B. Svetovoy, New. J. Phys. {\bf 8}, 238, (2006).
\bibitem{Casimir} H.G.B. Casimir, Proc. Kon. Ned. Akad. Wetenshap., {\bf B51}, 793, (1948).
\bibitem{Sparnaay} M.J. Sparnaay, in {\it Physics in the Making}, eds A. Sarlemijn, M.J. Sparnaay (North-Holland), 235 and references therein, (1989).
\bibitem{Milonni} P.W. Milonni, {\it The quantum vacuum}, Academic, (1994).
\bibitem{Mostepanenko} V.M. Mostepanenko, N.N. Trunov, {\it The Casimir effect and its applications}, Clarendon, (1997).
\bibitem{Lamoreaux} S.K. Lamoreaux, Resource Letter in Am. J. Phys. {\bf 67}, 890, (1999).
\bibitem{Bordag} M. Bordag, U. Mohideen and V.M. Mostepanenko, Phys. Reports {\bf 353}, 1, (2001) and references therein.
\bibitem{Reynaud} S. Reynaud, A. Lambrecht, C. Genet, M.T. Jaekel, C. R. Acad. Sci. Paris IV-2, 1287, and references
therein (2001).
\bibitem{Genet} C. Genet, A. Lambrecht, S. Reynaud, in {\it On the nature of dark energy}, eds. P. Brax, J. Martin, J.P. Uzan, 121, (Frontier Group, 2002). arXiv:quant-ph/0210173.
\bibitem{CP} H.G.B. Casimir, D. Polder, Phys. Rev. {\bf 73}, 360, (1948).
\bibitem{Babb} J.F. Babb, G.L. Klimchitskaya, V.M. Mostepanenko, Phys. Rev. {\bf A70}, 042901, (2004).
\bibitem{Antezza} M. Antezza, L.P. Pitaevskii, S. Stringari, Phys. Rev. Lett. {\bf 95}, 113202, (2005).
\bibitem{Long} J.C. Long, H.W. Chan, J.C. Price, Nucl. Phys. {\bf B539}, 23, (1999).
\bibitem{Adelberger} E.G. Adelberger, B.R. Heckel, A.E. Nelson, Ann. Rev. Nucl. Part. Sci. {\bf 53}, 77, (2003). arXiv:hep-ph/0307284.
\bibitem{comm_limits} The discrepancy between figures \ref{limits_short} and \ref{limits_long} around 1 micron can be traced back to the differing limits derived in \cite{Long} and \cite{Bordag2} from the same experiment \cite{Lamoreaux}. According to \cite{Adelberger} limits in figure \ref{limits_short} are less rigorous than those in figure \ref{limits_long}.
\bibitem{Bordag2} M. Bordag, B. Geyer, G.L. Klimchitskaya, V.M. Mostepanenko, Phys. Rev. {\bf D58}, 075003, (1998).
\bibitem{Courtillot} I. Courtillot, et al., Eur. Phys. J. {\bf D33}, 161, (2005).

\end{thebibliography}

\end{document}